\documentclass[pra,twocolumn,showpacs,groupedaddress]{revtex4}

\usepackage[dvips]{graphicx}
\usepackage{color}
\usepackage{latexsym}
\usepackage{amsmath}
\usepackage{amssymb}

\newcommand{\expect}[1]{\langle \Phi | #1 | \Phi \rangle}

\newcommand{\Ad}[1]{{\mathbf a}^\dag_{#1} }
\newcommand{\A}[1]{{\mathbf a}_{#1} }

\begin{document}

\preprint{LA-UR-05-2040}

\title
   {Acoustic attenuation rate in the Fermi-Bose model
    with a finite-range fermion-fermion interaction}
\author{Bogdan Mihaila}
\email{bmihaila@lanl.gov}
\affiliation{Theoretical Division,
   Los Alamos National Laboratory,
   Los Alamos, NM 87545}
\affiliation{Materials Science and Technology Division,
   Los Alamos National Laboratory,
   Los Alamos, NM 87545}

\author{Sergio~Gaudio}
\affiliation{Department of Physics,
   Boston College,
   Chestnut Hill, MA 02167}
\affiliation{Theoretical Division,
   Los Alamos National Laboratory,
   Los Alamos, NM 87545}

\author{Kevin~S.~Bedell}
\affiliation{Department of Physics,
   Boston College,
   Chestnut Hill, MA 02167}

\author{Krastan~B.~Blagoev}
\affiliation{Theoretical Division,
   Los Alamos National Laboratory,
   Los Alamos, NM 87545}

\author{Eddy Timmermans}
\affiliation{Theoretical Division,
   Los Alamos National Laboratory,
   Los Alamos, NM 87545}


\begin{abstract}
   We study the acoustic attenuation rate in the Fermi-Bose model
   describing a mixtures of bosonic and fermionic atom gases.
   We demonstrate
   the dramatic change of the acoustic attenuation rate as the fermionic
   component is evolved through the BEC-BCS crossover, in the
   context of a mean-field model applied to a finite-range
   fermion-fermion interaction at zero temperature, such as
   discussed previously by M.M.~Parish \emph{et al.}
   [Phys.\ Rev.\ B \textbf{71}, 064513 (2005)]
   and B.~Mihaila \emph{et al.}
   [Phys.\ Rev.\ Lett.\ \textbf{95}, 090402 (2005)].
   The shape of the acoustic
   attenuation rate as a function of the boson energy represents a
   signature for superfluidity in the fermionic component.
\end{abstract}

\maketitle


\section{Introduction}

Recent realization of superfluidity in a dilute gas of ultracold
fermionic atoms, is driving renewed theoretical efforts aiming at a
quantitative understanding of many-body correlations in a dilute
Fermi gas. At low density, for finite-range interactions, one
expects that all sensitivity to the detailed features of the
interaction is lost~\cite{schiff}, and the system's quantum behavior
depends only on the scattering length and the range of the
interaction. Furthermore, as we approach the ``unitarity''
limit~\cite{unitarity}, i.e. the limit where the scattering length
is much larger than the mean inter-particle separation, the system
exhibits universal behavior and the energy is determined entirely by
the density, leading to high sensitivity to many-body correlation
effects.

Built on the successful realization of quantum degenerate systems
consisting of boson~\cite{bosons} and/or fermion~\cite{fermions}
atoms, experimental progress in cold atom physics has recently
resulted in the generation of ultracold molecules from ultracold
atoms~\cite{creation}, Bose-Einstein condensates (BEC) of
molecules~\cite{molecules}, and the first experimental realization
of the small pair to large pair crossover~\cite{K40,Li6,NaLi}. In
one type of experiments~\cite{K40,Li6,NaLi}, the system consists of
one-type of cold, degenerate, fermionic atoms in an optical trap.
The strength of the interatomic interaction can be continuously
adjusted via a magnetically-tuned Feshbach resonance between a
closed and an open channel. Hence, the system can evolve from a BEC
phase in which spatially non-overlapping (Shafroth)
pairs~\cite{small-pairs} are bound together, to a
Bardeen-Cooper-Schrieffer (BCS)-type superfluid~\cite{bcs} involving
correlated atom pairs in momentum space.

Significant experimental work has been also directed towards the
study of mixtures of bosonic and fermionic atom
gases~\cite{stan,RbK}. Such mixtures remind of the strongly
correlated $^3$He-$^4$He mixtures~\cite{he3-he4}, and can be
obtained by sympathetically cooling a gas of fermionic atoms by
using a condensed boson gas as a refrigerator. In this paper, we
will focuss on the latter type of system, and discuss the fermionic
response to a density fluctuation of the Bose condensate in the
trap, with special emphasis on the effects due to the finite-range
character of the fermion-fermion interaction. We will argue that the
shape of the acoustic attenuation rate depends on the BEC or BCS
character of the fermionic liquid.


\section{Acoustic attenuation in the Fermi-Bose model}

The Hamiltonian describing the interacting Fermi-Bose system is
given as the sum of a boson, fermion and interaction components
\begin{equation}
   \mathbf{H}_{\mathrm {FB}}
   \ = \
   \mathbf{H}_{\mathrm B} \ + \
   \mathbf{H}_{\mathrm F} \ + \
   \mathbf{H}_{\mathrm{int}}
   \>.
\end{equation}
This model, proposed in this context by Timmermans \emph{et
al}~\cite{fermibose}, is similar to the a model introduced by
Bardeen \emph{et al}~\cite{bardeen} to describe the effect of the
fluctuations of the polarization in an excitonic liquid and a gas of
electrons. This model has since been the subject of intense
investigations in connection with the Feshbach resonance mechanism
in ultracold fermionic gases~\cite{griffin,holland}.



In the weak coupling limit, the Fermi and Bose components can be
treated independently, and the many-body wave function of the
Fermi-Bose mixture is the direct product of \emph{independent} Fermi
and Bose wave functions, $| \Phi_{FB} \rangle = | \Phi_{B}, \Phi_{F}
\rangle$. The dispersion relation of the dilute Bose condensate
is~\cite{eddy_cp}
\begin{equation}
   \omega_q \ = \ c \, q \, \sqrt{1 + (\xi_B \, q)^2}
   \>,
\label{eq:Eq}
\end{equation}
where $c = \sqrt{\lambda_B \, \rho_B / m_B}$ is the phonon (BEC
excitation) velocity of sound, $\xi_B = 1/\sqrt{4\lambda_B \rho_B
m_B}$ is the boson coherence length, and $\lambda_B = 4\pi a_B/m_B$
is the strength of the bosonic pseudo-potential. The coherence
length, i.e. the length scale at which the condensate varies
spatially, is typically of the order of $\xi_B \sim 0.1 - 1\,
\rm{\mu m}$, while $c$ is approximately $0.01~\rm{\mu m\, MHz}$ in
condensates with density of $100\, \rm{\mu m}^{-3}$. The scattering
length, $a_B$, and the density of the bosonic component, $\rho_B$,
are tunable parameters which can be used to modify the dispersion
relation~(\ref{eq:Eq}), but $c \, \xi_B = 1 / (2\, m_B)$, $c/\xi_B =
2 \, \lambda_B \, \rho_B \sim \rho_B \, a_B$ and, therefore, the
parameters $c$ and $\xi_B$ cannot be varied independently.
Throughout this paper $\hbar=1$, i.e. the energy is measured in
units of frequency.



The low-energy state of the system is described by the free-energy
associated with the BCS-reduced Hamiltonian
\begin{align}
   \mathbf{H}_{\mathrm F} \ = \ &
   \sum_{\mathbf{k}} \ \epsilon_{\mathbf{k}} \,
   \bigl ( a^{\dag}_{\mathbf{k} \uparrow} a_{\mathbf{k} \uparrow}
   +
   a^{\dag}_{\mathbf{k} \downarrow} a_{\mathbf{k} \downarrow}
   \bigr )
   \\ \notag &
   +
   \sum_{\mathbf{k,p,q}} V_{\mathbf{q}} \
                 a^{\dag}_{\mathbf{k} \uparrow}
                 a^{\dag}_{\mathbf{p} \downarrow}
                 a_{\mathbf{p} - \mathbf{q} \downarrow}
                 a_{\mathbf{k} + \mathbf{q} \uparrow}
   \>,
\label{eq:ham_0}
\end{align}
with $\epsilon_{\mathbf{k}} = \mathbf{k}^2 / (2m_F)$. Then, in a
zero-temperature mean-field
approximation~\cite{ref:leg80,ref:CN82,2level}, the fermionic
ground-state in the Hartree-Fock-Bogoliubov representation~\cite{BV}
is the standard BCS variational wave function given by
\begin{equation}
   | \Phi_F \rangle \ = \
   \mathcal{N} \
   \exp \left ( \sum_{\mathbf{k}} \ \frac{v_{\mathbf{k}}}{u_{\mathbf{k}}} \
   a^{\dag}_{\mathbf{k} \uparrow} a^{\dag}_{-\mathbf{k} \downarrow}
   \right ) \ | 0 \rangle
   \>.
\end{equation}
The BCS ansatz interpolates smoothly between the BCS and BEC limit.
The parameters $\{ u_{\mathbf{k}}, \, v_{\mathbf{k}} \}$ are
obtained by solving the system of equations $
   v_{\mathbf{k}}^2
   =
   \frac{1}{2} -
   (\epsilon_{\mathbf{k}} - \mu)/(2 E_{\mathbf{k}})
$ and
$
   u_{\mathbf{k}} v_{\mathbf{k}}
   =
   \Delta_{\mathbf{k}}/(2 E_{\mathbf{k}})
$, subject to the normalization condition, $| u_{\mathbf{k}}|^2 +
|v_{\mathbf{k}}|^2 = $~$1$. Here, $\mu$ is the fermionic chemical
potential such that spin-up and spin-down states are equally
occupied, and $E_{\mathbf{k}}$ denotes the quasi-particle spectrum
in the fermionic ground state, $E_{\mathbf{k}}^2 =
(\epsilon_{\mathbf{k}} - \mu)^2 + \Delta_{\mathbf{k}}^2$. The
pairing gap, $\Delta_{\mathbf{k}}$, is obtained as
$
   \Delta_{\mathbf{k}}
   =
   -
   \sum_{\mathbf{p}}
   V_{\mathbf{k}-\mathbf{p}} \,
   \Delta_{\mathbf{p}}/E_{\mathbf{p}}
$.



The interaction between the Fermi and Bose components is assumed to
be weak, and the interaction Hamiltonian,
$\mathbf{H}_{\mathrm{int}}$, describing the interaction of a single
acoustic phonon of momentum $\mathbf{q}$ and energy $\omega_q$ with
the fermionic component can be written similarly to the
electron-phonon Hamiltonian~\cite{mahan}, as:
\begin{equation}
   \mathbf{H}_{\mathrm{int}} \sim
   \sqrt{\frac{1}{2 \rho \, \omega_q}} \
   \int \!\! \frac{\mathrm{d}^3 k}{(2\pi)^3} \
   | \mathbf{q} | \
   \mathbf{b}^\dag_{\mathbf{q}} \,
   \sum_{\sigma=\uparrow\!, \downarrow}
   \mathbf{a}^\dag_{\mathbf{k}-\mathbf{q}, \sigma} \,
   \mathbf{a}_{\mathbf{k} \sigma}
   \ + \ \mathrm{h.c.}
   \>,
\end{equation}
where $\rho$ is the mass density of the composite gas. (For an
alternate approach based on a density-density interaction, see
Ref.~\cite{paper_one_att}.) In first-order perturbation
theory~\cite{first}, the acoustic attenuation rate is given by
\begin{equation}
   \Gamma(q) \ = \
   2\pi \
   |\, \langle i | \mathbf{H}_{\mathrm{int}} | f \rangle \, |^2 \
   \delta(E_i - E_f)
   \>.
\end{equation}
At zero temperature, for a spin-unpolarized Fermi gas, we
obtain~\cite{mahan}:
\begin{align}
   & \Gamma(q) \sim
   \frac{q^2}{\omega_q} \
   \bigl [ \gamma_{\rho}(q) + \gamma_{\kappa}(q) \bigr ]
\label{integral}
   \\ \notag & \sim
   \frac{q^2}{\omega_q} \!
   \int \!\! \frac{\mathrm{d}^3 k}{(2\pi)^3}
   \bigl [ \rho_\mathbf{k} (1 - \rho_{\mathbf{k}'})
           + \kappa_\mathbf{k} \kappa_{\mathbf{k}'}
   \bigr ]
   \delta(\omega_q - E_k - E_{k'})
   \>,
\end{align}
with $\mathbf{k}'=\mathbf{q}+\mathbf{k}$. In the mean-field
approximation, the normal and anomalous densities in the fermionic
ground state are defined as $\rho_{\mathbf{k}} = \expect{
\Ad{\mathbf{k} \uparrow} \A{\mathbf{k} \uparrow} } =
|v_{\mathbf{k}}|^2$, and $\kappa_{\mathbf{k}} = \expect{
\A{-\mathbf{k} \downarrow} \A{\mathbf{k} \uparrow} } =
v_{\mathbf{k}}^* u_{\mathbf{k}}$. The acoustic attenuation rate
appears as the superposition of components $\gamma_{\rho}$ and
$\gamma_{\kappa}$, which depend separately on the normal and
anomalous densities, respectively (see also Ref.~\cite{response}).

In principle, one should also consider the possibility of exciting
the Anderson-Goldstone mode of the fermion superfluid system.
However, as it has been discussed in Ref.~\cite{paper_one_att}, the
Anderson-mode excitation can be avoided by specifying that the BEC
velocity of sound, $c$, is less than the velocity of the Anderson
mode, $v$, i.e. we require $c < v$. Then, the excitation of the
collective Anderson mode by a BEC phonon mode is energetically
forbidden. This is a reasonable assumption for the mixtures of
bosonic BEC and fermionic atom gases under consideration here: In
the BCS regime, the velocity of the Anderson mode, $v$, is equal to
the Fermi-velocity $v_F$ and, for typical atom trap conditions, that
can be easily satisfied. The only cold atom mixtures for which $c >
v_F$ contain ultra-low density fermion systems or ultra-high density
BEC's: the first type would be difficult to realize because of the
very low critical temperature for fermion superfluidity, the second
type would be difficult to achieve experimentally because three-body
recombination processes would quickly deplete the experimental
system. Hence, while the Anderson mode will most likely be somewhat
softened in the crossover regime (since we expect it to go over into
the Bogoliubov mode of the BEC-system in the BEC limit of the
crossover), by choosing a sufficiently weakly interacting BEC or low
density BEC, it should always be possible to ensure that $c < v$
throughout the entire crossover.


\section{A simple model}

In order to study the phase-space constraints subjecting the
acoustic attenuation rate, we consider first a schematic model of
the momentum-dependence of the pairing gap, suggested by Comte and
Nozi\`eres~\cite{ref:CN82}:
\begin{equation}
   \Delta_k \ = \ \frac{\Delta_0}{1 + (k/k_F)^2}
   \>.
   \label{eq:CN}
\end{equation}
Note that the realistic momentum dependence of the pairing gap, as
predicted by the one-channel fermion-fermion model with a
finite-range interaction to be discussed in the next section, has a
much longer tail than the simple ansatz of Comte and
Nozi\`eres~\cite{ref:CN82}. Unless otherwise specified, in the
following we set $\hbar c k_F/\varepsilon_F=1$ for simplicity. We
also keep fixed the chemical potential, $\mu/\varepsilon_F$, and the
coherence length of the Bose-gas component, $\xi_B k_F$, and vary
the pairing-gap parameter~$\Delta_0/\varepsilon_F$.


\begin{figure}[t!]
   \includegraphics[width=0.95\columnwidth]{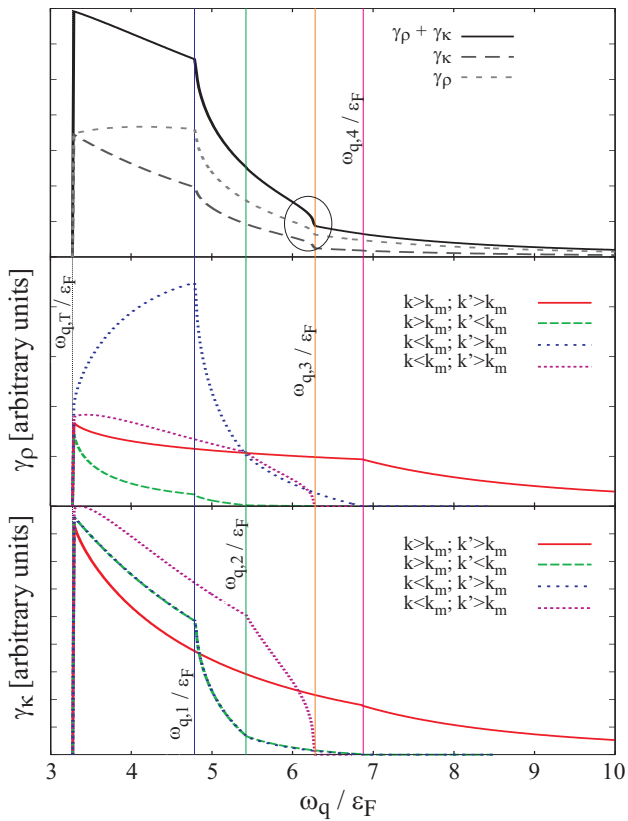}
   \caption{\label{fig:D4} (Color online)
   Acoustic attenuation rate components
   corresponding to the toy model described by Eq.~\eqref{eq:CN}.
   Here we study a large pairing-gap, $\Delta_0/\varepsilon_F$, regime:
   we set $\Delta_0/\varepsilon_F=4,\ \mu/\varepsilon_F = 1$,
   and $\xi_B k_F=5$.
}
\end{figure}


\begin{figure}[t!]
   \includegraphics[width=0.95\columnwidth]{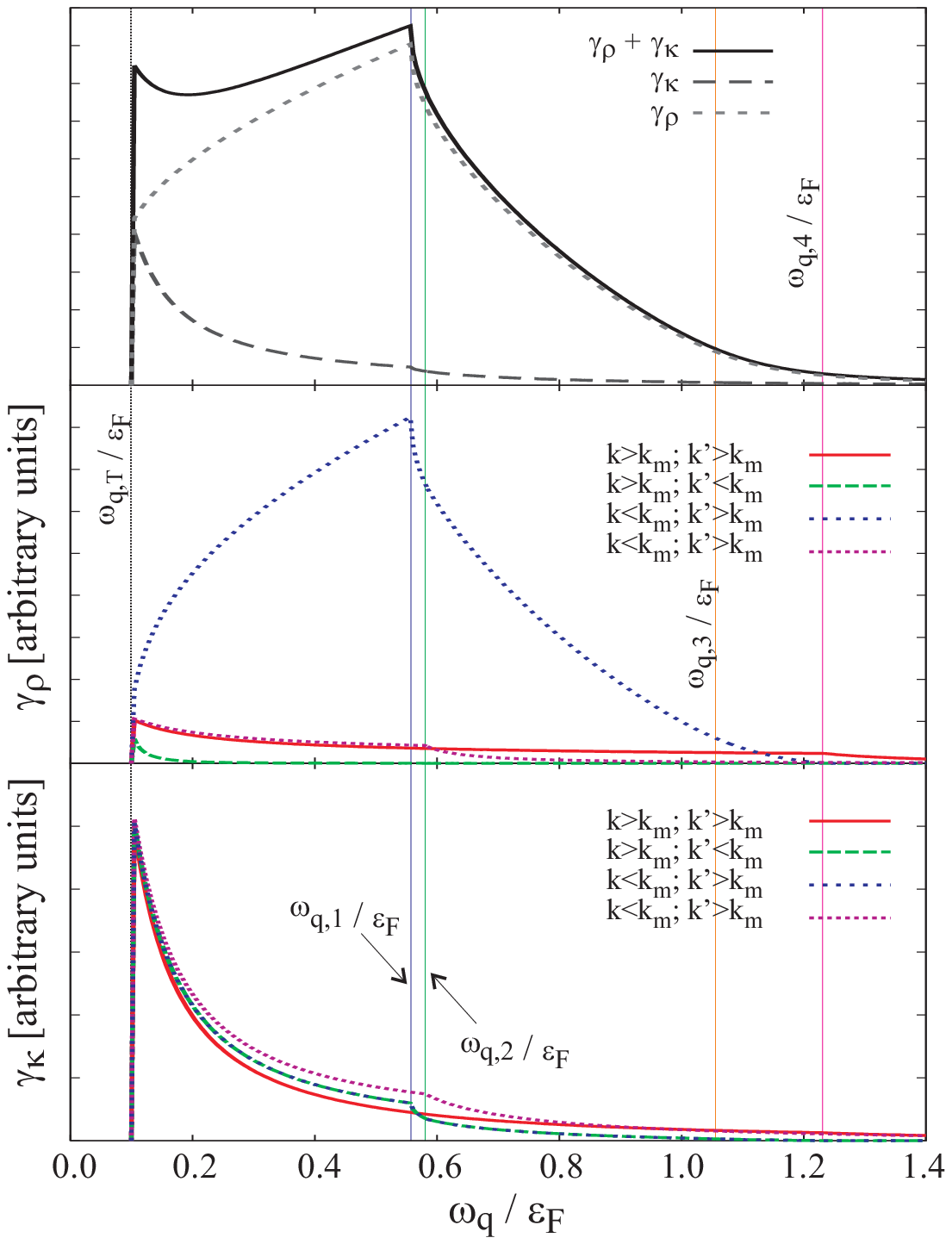}
   \caption{\label{fig:D01} (Color online)
   Acoustic attenuation rate components
   corresponding to the toy model described by Eq.~\eqref{eq:CN}.
   Here we study a small pairing-gap, $\Delta_0/\varepsilon_F$, regime (BCS-like):
   we set $\Delta_0/\varepsilon_F=0.1,\ \mu/\varepsilon_F = 1$,
   and $\xi_B k_F=5$.
}
\end{figure}


\begin{figure}
   \includegraphics[width=0.95\columnwidth]{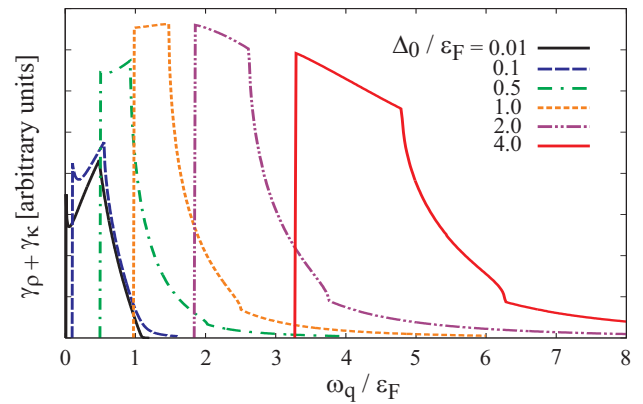}
   \caption{\label{fig:Eb} (Color online)
   Shapes of the energy-dependence of the attenuation-rate
   as a function of the pairing-gap
   parameter, $\Delta_0/\varepsilon_F$, for fixed values
   of $\mu/\varepsilon_F = 1$ and $\xi_B k_F=5$.
}
\end{figure}


For illustrative purposes, in Figs.~\ref{fig:D4} and~\ref{fig:D01},
we plot the energy-dependence of the various attenuation-rate
components, $\gamma_{\kappa}$ and $\gamma_{\rho}$, for the regime of
large and small values of the pairing-gap, $\Delta_0/\varepsilon_F$,
respectively. Because the quasi-particle spectrum in the model
described by Eq.~(\ref{eq:CN}) has a minimum at finite momentum,
$k_m$, then the conservation of energy, $E_{k'}=\omega_q - E_k$, and
momentum, $\mathbf{k}' = \mathbf{q} + \mathbf{k}$, leads to four
branches. The available phase-space satisfying these constraints
leads to kinks observed in the energy-dependence of the acoustic
attenuation rate. In particular, we have four ``critical" points, as
depicted by the vertical lines in Figs.~\ref{fig:D4}
and~\ref{fig:D01}. In the increasing order of the ``critical''
energy, we have: $\omega_{q,1}$~the energy value at which the
momentum conservation corresponding to the second ($k>k_m,\ k'<k_m$)
and third ($k<k_m,\ k'>k_m$) branches, is no longer satisfied. This
failure generates the most remarkable feature in the acoustic
attenuation rate, resulting in a decrease in the acoustic
attenuation rate; $\omega_{q,2}$~the energy value at which the
fourth branch ($k<k_m$, $k'<k_m$) also fails due to failed momentum
conservation; $\omega_{q,3}$~the energy value at which the fourth
branch vanishes; and $\omega_{q,4}$~the energy value at which the
second and third branches also vanish: from this point on, the only
contribution is from the first branch ($k>k_m,\ k'>k_m$).

The energy distribution of the acoustic attenuation rate may feature
a ``shoulder'' just before $\omega_{q,3}$, because the domain of the
integrals corresponding to branches three and four is $k \in
[0,k_m]$, i.e. the integration domain for these branches is
$\omega_q$-independent (see circled area in the top panel of
Fig.~\ref{fig:D4}. This ``shoulder'' is apparent in the large
$\Delta_0/\varepsilon_F$ regime, and disappears as we approach the
normal Fermi gas limit.


In the small $\Delta_0/\varepsilon_F$ regime, the ``critical''
energies $\omega_{q,1}$ and $\omega_{q,2}$ are close together, while
the ``critical'' energies $\omega_{q,3}$ and $\omega_{q,4}$ are
pushed to relatively higher values. When $\Delta_0/\varepsilon_F$
becomes very small, only one of the eight contributions to $\Gamma$
is significant (due to the fact that the anomalous density is a
delta-like distribution centered around~$k_F$), and the kinks in the
energy distribution of the acoustic attenuation rate are entirely
obscured, except for~$\omega_{q,1}$.

In Fig.~\ref{fig:Eb} we show evolution of the attenuation rate
profile as a function of the pairing-gap, $\Delta_0/\varepsilon_F$,
at fixed chemical potential. The shape of the attenuation rate
changes smoothly between low and large values of the pairing-gap.
The signature of the crossover is the change in the sign of the
slope of the acoustic attenuation rate, for phonon energies between
the threshold energy, $\omega_{q,T} = 2 E_{k_{\rm{m}}}$, and the
first ``critical'' value, $\omega_{q,1}$: the slope is positive in
the small $\Delta_0/\varepsilon_F$ regime, and becomes negative in
the large $\Delta_0/\varepsilon_F$ regime. From Figs.~\ref{fig:D4}
and~\ref{fig:D01} follows that , the crossover is characterized by
the ratio of $\gamma_{\kappa}/\gamma_{\rho}$, i.e. the ratio of the
anomalous- to the normal-density. In the small
$\Delta_0/\varepsilon_F$ limit this ratio is small (it vanishes for
the Fermi gas). In the large $\Delta_0/\varepsilon_F$ limit, the
anomalous-density contribution is comparable in size (but still
smaller) than the normal-density contribution. In the limit when
$\Delta_0\rightarrow 0$, we recover the normal Fermi gas result.


\section{Effects of a finite-range fermion-fermion interaction}

We consider now a more realistic scenario, in which the fermionic
atoms interacting via a short-range Gaussian potential, $V(r) = V_0
\exp(- b r^2)$. The mean-field approximation for the ground-state
properties of this system has been discussed recently in
Ref.~\cite{2level}, where the BEC-BCS crossover was studied by
fixing the width of the potential $\langle r \rangle = 2/\sqrt{\pi
b}$ and varying the potential depth $V_0$, or implicitly the
scattering length~$a_F$ of the interaction. Following
Leggett~\cite{ref:leg80}, the dilute fermionic state is
characterized by the parameter $\eta = (k_F a_F)^{-1}$, which
combines together the density of the fermionic gas, characterized by
the Fermi momentum, $k_F$, and the fermion-fermion scattering
length. The BEC-BCS crossover is indicated by the disappearance of
the singularity in the fermionic density of states,
$
   N(k)
   \sim
   k^2 
   \left | \frac{\mathrm{d}E_k}{\mathrm{d}k} \right |^{-1}
$ (see Fig.~3 in Ref.~\cite{2level}). These results are similar to a
magnetic field dependent contact interaction~\cite{marzena}, and are
relevant to present experiments in which only the lowest hyperfine
levels of the fermionic atom gas are populated.


\begin{figure}[t]
   \includegraphics[width=0.95\columnwidth]{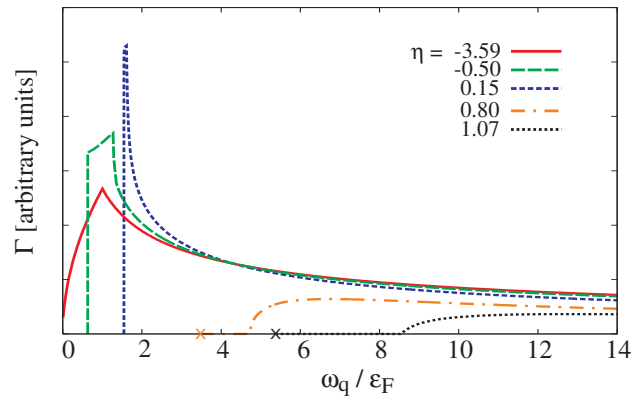}
   \caption{\label{fig:one_att} (Color online)
   Energy dependence of the acoustic attenuation rate as the
   fermionic component goes through the BEC-BCS crossover,
   for the choice of parameters in Ref.~\cite{2level}.
   Along the $\omega_q$-axis, we plot the position of the energy
   thresholds, $\omega_{q,T} = 2 E_{k_{\rm{m}}}$.
   }
\end{figure}


\begin{figure}
   \includegraphics[width=0.95\columnwidth]{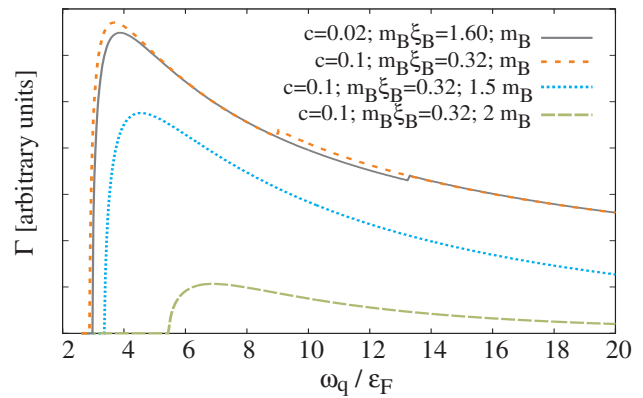}
   \caption{\label{fig:alpha} (Color online)
   Effect of the bosonic parameters
   on intensity of the BEC signal.
   }
\end{figure}


To study the effect of the energy-momentum dependence of the normal
and anomalous densities on the acoustic attenuation rate, we
consider the BEC-BCS crossover at constant-density corresponding to
$k_F \langle r \rangle \approx~0.37$. The BEC-BCS crossover in this
case occurs for $\eta = (k_F a_F)^{-1}$ between 0.15 and 0.80. We
assume that the dispersion relation of the bosonic component is
given by the parameters: $c=0.02\, \rm{\mu m\, MHz}$, and $m_B \xi_B
= 1.6\, (\rm{u})(\rm{\mu m})$. Our results are illustrated in
Fig.~\ref{fig:one_att}. Here, the energy dependence of the
attenuation rate, $\Gamma$, as a function of the scattering length
(or $\eta = (k_F a_F)^{-1}$), is plotted at fixed fermion density
($k_F \langle r \rangle \approx~0.37$). For our choice of parameters
in the dispersion relation of the dilute Bose condensate (see
Eq.~\eqref{eq:Eq}), the ratio $q^2/\omega_q$ approaches a constant
for $\omega_q/\varepsilon_F \ll 1$. Here $k_F$ is the Fermi
momentum, and $a_F$ is the s-wave scattering length in the fermionic
component.

At the threshold, on the BCS side, the acoustic attenuation rate
features a sharp edge, followed by a (linear) increase in the
acoustic attenuation rate as a function of the phonon energy; in
contrast, on the BEC side, the sharp edge is smoothed out, and the
acoustic attenuation rate decreases slowly as a function of energy.
It is important to recall that in the dilute limit the BEC-BCS
crossover coincides with the singularity in the scattering length,
and occurs for $\mu=0$~\cite{ref:leg80}. The acoustic attenuation
rate provides a crisp signature of the BEC-BCS crossover.


On the BEC side, the energy threshold in first-order perturbation
theory is higher than the energy threshold ($\omega_{q,T} = 2
E_{k_{\rm{m}}}$) given by energy conservation, due to the reduced
phase space availability for momentum conservation. In this regime
we have $k_m=0$ and the contributions due to the third and fourth
branches ($k<k_m$) vanish. Due to the lack of available phase space
for the integral~\eqref{integral}, the actual threshold in
first-order perturbation theory is located further away from the
energy threshold as we go deeper into the BEC regime (see
Fig.~\ref{fig:one_att}). The effect is enhanced by a smaller bosonic
coherence length.

From the numerical results depicted in Fig.~\ref{fig:one_att} it is
apparent that the attenuation rate calculated in first-order
perturbation theory vanishes in the deep BEC region. In this regime,
one needs to include a second-order approximation in order to
describe the scattering of the boson off the molecular condensate.
Provided that i)~the molecular condensate degrees of freedom do not
play a role and that ii)~the coupling between the bosonic and
fermionic components is sufficiently weak this second-order
correction in perturbation theory is expected to be small and will
not be discussed here. For energies greater than the threshold, the
acoustic attenuation rate becomes nonzero and the first-order result
is expected to be dominant.

The dependence on the phonon velocity and/or the boson mass of the
strength of the BEC signal is illustrated in Fig.~\ref{fig:alpha}.
Here, we depict the acoustic attenuation rate for $\eta=0.59$ (the
crossover region), as obtained by first changing the phonon velocity
from 0.02 to $0.1\, \rm{\mu m\, MHz}$, which is equivalent to
changing $m_B \xi_B$ from 1.6 to $0.32\, (\rm{u})(\rm{\mu m})$.
Next, we increase the boson mass, $m_B$, first by 50~\% and then by
100~\%, which is equivalent to reducing by the same factor the
bosonic coherence length, $\xi_B$. The acoustic attenuation rate
changes little when we increase the phonon velocity, $c$, for a
given boson mass, $m_B$. In contrast, for heavier bosons, the signal
drops in intensity very quickly for a relatively small increase of
the boson mass, or else for a relatively small decrease in the
bosonic coherence length.

We note that in a related formalism, similar shapes (especially on
the BEC side) are present in the discussion of the spin-spin
correlation functions (see Refs.~\cite{Buchler,response}). These
calculations refer to observables accessible via Bragg, rf and
spin-noise spectroscopy. Therefore, measurements of the acoustic
attenuation rate can be used in conjunction with these other
techniques to probe features of the fermionic normal and anomalous
densities. The acoustic attenuation rate discussed here is
intrinsically a nonzero momentum transfer measurement, thus probing
different moments of the fermionic normal and anomalous densities.
This in turn results in different features in the shapes on the BCS
side.


\section{Conclusions}

To summarize, in this paper we have studied the acoustic attenuation
rate in the framework of the Fermi-Bose model, for a realistic
finite-range fermion-fermion interaction. Our findings suggest that
by inducing a density fluctuation in the Bose gas component one can
induce an acoustic response in the fermionic component, and the
shape of the acoustic attenuation rate can be used as a signature of
superfluidity on the BCS side. For a low excitation energy, the
acoustic attenuation rate increases with the phonon energy on the
BCS side and decreases on the BEC side. Therefore, a linear increase
of the acoustic attenuation rate with the phonon energy above the
threshold indicates that one has passed to the BCS side. Moreover,
our results show that the signal on the BCS side is much larger than
on the BEC side, which makes the detection easier. The acoustic
attenuation rate measurement is favored by a larger phonon velocity
and the choice of the lightest available boson specie.


\begin{acknowledgments}
This work was supported in part by the LDRD program at Los Alamos
National Laboratory. B.M. acknowledges financial support from an
ICAM fellowship program. The authors would like to thank
M.M.~Parish, P.B.~Littlewood, and D.L.~Smith for useful discussions.
\end{acknowledgments}


\vfill

%
%


\begin{thebibliography}{99}
\bibitem{first}
Perturbation theory with quasi-particle states is perfectly valid
even if the system is strongly interacting: quasi-particle states
interact weakly with each other, even if the underlying (real)
fermions of the system interact strongly. Here, the expectation
approach is that the quasi-particle picture remains valid for
low-energy fermion modes even if the fermions become strongly
interacting. This is the case for normal fermion systems, and this
phenomenon, in which the Pauli-principle plays a major role, forms
the basis of the Landau description of Fermi-liquid theory. In this
case, a single low-energy fermion quasi-particle state near the gap
could remain long lived as well, for reasons of kinematics and
fermion statistics.
\end{thebibliography}
\end{document}